\documentclass[runningheads,a4paper]{llncs}
\usepackage{url}
\usepackage{amsmath}
\usepackage{listings}
\usepackage{times}
\usepackage{helvet} 
\usepackage{courier}
\usepackage{graphicx}
\usepackage{caption}
\DeclareCaptionType{copyrightbox}
\usepackage{subcaption}
\usepackage{verbatim}

\usepackage{ifpdf}
\ifpdf
\else
  
\fi
\usepackage[margin=1.25in]{geometry}

\urldef{\mailsa}\path|{pathania.deepika@gmail.com, kamal@iiit.ac.in}|
\newcommand{\keywords}[1]{\par\addvspace\baselineskip
\noindent\keywordname\enspace\ignorespaces#1}

\begin{document}

\mainmatter  

\title{Exploiting Near Time Forecasting From Social Network To Decongest Traffic}

%
%
\author{Deepika Pathania \and Kamalakar Karlapalem}
%

\institute{Center for Data Engineering,\\
International Institute of Information Technology Hyderabad, India\\
\mailsa\\}

%
%

\maketitle

\begin{abstract}
Preventing traffic congestion by forecasting near time traffic flows is an important problem as it leads to 
effective use of transport resources. Social network provides information about activities of humans and social events. 
Thus, with the help of social network, we can extract which humans will attend a particular event (in near time) and can estimate 
flow of traffic based on it. This opens up a wide area of research which poses need to have a framework for traffic management that can 
capture essential parameters of real-life behaviour and provide a way to iterate upon and evaluate 
new ideas. In this paper, we present building blocks of a framework and a system to simulate a city with its transport system, 
humans and their social network. We emphasize on relevant parameters selected and modular design of the framework. 
Our framework defines metrics to evaluate congestion avoidance strategies. To show utility of the framework, 
we present experimental studies of few strategies on a public transport system.
\end{abstract}








\section{Introduction}
Traffic congestion is a major problem that is faced in many parts of the world.
It wastes two precious resources - time and fuel. 
According to Texas A\&M Transportation Institute's 2012 Urban Mobility Report \cite{texas:report}, congestion cost of extra time and fuel is \$121 billion in monetary terms 
for 498 urban areas in 2011; and for an average commuter, it costs more than \$2 per day.
This cost has increased from \$40 billion to \$121 billion over last 3 decades. 
There can be many causes responsible for congestion: (i) traffic 
(demand) more than road-rail capacity (supply); for example, increase in ridership of public transportation like trains due to social events (ii) unexpected human movement activities, for example, vehicle breakdowns on roads/rails, events like concerts in a city.
Public transport systems are usually prepared for routine 
movement of humans, but not for unexpected events; either they do not have prior information (accidents, etc.) or not enough information (like an estimate on number of humans that may visit a concert and which routes would see an increased use).

Traffic prediction is a possible solution to solve congestion problem.
If we have prior knowledge about extent of traffic at particular places, we can 
make arrangements beforehand and avoid congestion; like having extra buses to meet increase in ridership.
Many solutions have been studied regarding prediction of traffic.
In real time sensing of traffic, current behavior and flow of traffic is judged (using sensors, image processors, etc.) and then congestion-prone areas are found. These methods provide information about likelihood of congestion at a short notice and it might not be feasible to control congestion formation.
Analyzing historical data related to traffic behavior is another approach to anticipate traffic congestions.
Aforementioned techniques do not provide movement plan and travel destination of humans.
If we can estimate travel plans well before, we know the immediate future flow of traffic and congestion avoidance plans can be implemented in a timely and efficient manner.

\subsection{Motivation}
With the emergence of social networking websites, we explore another approach for traffic prediction.
Nowadays, many humans update their daily activities or plans on social websites. 
Further, these websites have also become a media of rapid information dissemination.
Humans notice events (like a concert or a spot-sale) and quickly spread this information to thousands of others.
Social network data can potentially (with some limitations due to privacy) enable us to estimate which humans will attend a particular event. 
We can estimate the flow of traffic using humans' source and destination locations.
This will allow us to estimate traffic flow well before time and thus give us more time to explore possibilities of actions that could be taken to control traffic congestion.
This problem domain opens up a large area of research.

In our work, we present a framework that simulates essential components of the problem:
\begin{enumerate}
  \item a city with its points of interests (homes, offices, train-stations, schools, theatres, hospitals, parks, etc.);
  \item a human population with their daily activities;
  \item a social network of humans;
  \item a transport management system that monitors traffic and plans to control congestion;
  \item social events which cause anomalies in traffic pattern.
\end{enumerate}
Our design and framework for solving the problem is generic and components are flexibly pluggable to be useful and extendible by other researchers.
We use Singapore city to explain framework parameters and present results of simulation.

We consider trains as means of transport in our framework.
Areas of congestion is limited to train stations and elements that form congestion are humans (and not any vehicles).
Trains take commuters from one train station to another.
There is a waiting period for humans at train stations.
If an already full train arrives, then humans will not be able to get into train and have to wait for the next train that can cause congestion.
If there is too much congestion at train station, it will take a long time before a human can board a train.
Hence, the transport management system aims to plan train services such that humans have to wait as less as possible.

We have used agent-based approach to develop our framework. 
Humans carry out their daily activities like going to schools, offices, attending social events, etc. 
Each human takes autonomous decisions in picking up road and rail route to reach a destination.
Human autonomously decides whether to attend a social event or not based on its interactions with its friends.
Transport system (section \ref{transport-system}) consists of trains, stations, a train inquiry system, a transport manager and one station master for each station.
Station master tracks ridership at its station and asks transport manager to take actions if ridership increases beyond expectation.
Additionally, it regulates arrival of trains at a station when number of trains waiting to arrive at station is more than number of available platforms.
Transport manager frames train schedule and routes such that waiting time of humans at stations is as less as possible. 
This is standard operations research problem and not part of this paper.

Agent-based architecture facilitates these individual behaviors to different actors in the simulation and manages interactions among them.
Our framework models a social network composed of humans and friend relationships amongst them.
Humans post updates on social network about their activities, interests and their plans (for example, attending some social events) and quickly spread information about events; these posts may in turn influence more humans.
Using these posts, the transport manager extracts information about events which are going to be held in near future - their locations and timings.
The transport manager also analyzes these posts to find out humans who plan to attend these events in order to estimate which regions of city are expected to face increase in demand of public transportation.
This approach enables transport manager to foresee human movement and take measures to meet increase in ridership of trains to avoid making humans wait for trains at stations.
This gives more reaction time than online prediction techniques suggested in \cite{Jain:image}, \cite{vaqar:traffic}, \cite{Ando:pheromone}, \cite{ben:dynamit}.
Since, the posts provide knowledge of travel plans of humans, it increases the probability of accuracy in traffic prediction at different stations.

This paper is organized as follows.
Section \ref{related-work} presents current state of related work.
Section \ref{event-broadcaster} presents a module that generates events with same temporal and geographical distribution as in real-life Singapore.
Section \ref{humans-social-network} presents how we model humans, social network and how a human influences its connection.
Section \ref{transport-system} presents the components of transport system.
Finally, we present results in section \ref{results} and conclude the paper in following section.

\section{Related Work}
\label{related-work}
Several models of simulations of traffic and transport systems have 
been studied and implemented focusing on finding the regions which could see traffic jams or congestion in near future.
We divide these approaches in two categories:
(i) traffic simulation; and (ii) traffic prediction. 

\subsection{Traffic Simulation}
TRANSIMS \cite{smith:transims} is an integrated set of tools developed to conduct regional transportation system analyses. 
This system models population with daily activities and their travel movements on a detailed represented transportation network.
SUMO \cite{kra:sumo} simulates road network traffic.
It models collision free vehicle movements on road and safe junctions with traffic lights.
DYNASMART \cite{hu:dynasmart} is a time-based microscopic approach for modeling traffic flow and
evaluating overall network performance under real-time information systems.
It simulates individual vehicles and each vehicle chooses its path from k-shortest paths according to its characteristics.
It can model traffic flow dynamics such as congestion formation and studying different control strategies.
It also provides route guidance, transport network planning and traffic operation decisions.
TraMas \cite{fernandes:tramas} is a multi-agent based approach in road traffic simulation.
It includes three layers - cooperative, decision and control.
In this system, each cross-section has a traffic agent to control traffic at that location by cooperating with other agents.
\cite{nagel:sim} presents a system consisting of mobile human agents whose travel behavior is modeled close to real life, i.e., strategies for human agents are activity scheduling, route choice, reaction to congestion and choice of residence or workplace. 

\subsection{Traffic Prediction}
Real-time sensing of traffic on road, statistical methods based 
on past history and simulation based techniques have been studied 
for prediction of traffic congestion. 
\subsubsection{Social network based approach}
\cite{social-network-china} presents a study of use of social media in finding solution to traffic congestion problem happening during golden week (Sep 30th - Oct 7th) 
in China. Golden week and travel were found as highly popular topics in social networking sites.
Based on analysis of travel related topics on social media for the month of September, authors present how on-line users' attributes and published content can be employed to forecast the geographic distribution of potential tourists.
Another example of investigating social media data to predict traffic 
congestion is studied in \cite{tweet-ijcai}. Authors extract tweet semantics and predict traffic congestion by incorporating semantics in auto-regression model.
STAR-CITY (Semantic Traffic Analytics and Reasoning for City) \cite{star-city} is an innovative system for diagnosis/prediction of traffic conditions and efficient urban 
planning which semantically analyzes heterogeneous data from various sources like historical data, present road/weather information, public transport 
stream, social media feeds and city events. Our work provides a framework for implementing different approaches for analyzing data of social media and extracting events; 
in addition it also provides a way of experimenting with strategies that can control congestion and improve traffic management.

\subsubsection{Conventional sensing approach}
This approach deals with keeping track of count of vehicles moving on segment of road and estimates their speed which is used for detection of congestion areas.
Conventional sensors like inductive loop detectors, magnetic sensors, video image processors, etc. are used for detection \cite{handbook:report}.
Image processing on images captured by traffic cameras is another example of detecting state of congestion \cite{Jain:image}. 
GPS technology has been used for tracking movements of vehicles on road \cite{vaqar:traffic}.
Vehicles equipped with GPS systems form a vehicular network.
They communicate within the network and collect information like speed, position and direction of moving from other participating vehicles.
This information helps to detect the state of congestion areas and further helps to avoid congestion.

Sensor based approaches require installation and maintenance of different equipments for sensing.
These approaches are suitable in predicting traffic of only near time giving transport management less time to take actions.

\subsubsection{Statistical approach}
Statistical methods of traffic prediction learns about the traffic conditions from the past traffic patterns.
Some of these methods are based on neural networks \cite{report:neural} and Bayesian networks \cite{yoshii:bayesian}. 
Clustering trajectories \cite{Gaffney:cluster}, \cite{Lee:cluster} is another 
way to find out routes with high traffic load. Regression analysis \cite{oswald:regression} 
has also been proposed to predict future motion of individuals.

\section{Event Broadcaster}
\label{event-broadcaster}
Social events like spot sale or a celebrity function can occur and are identified at short notice before their actual occurrence. 
Their information is published on websites,
in newspapers, popularized on social networks, etc.
Humans interact with these media on regular basis and come to know about such events.
We encapsulate these media into a single module called \textit{event broadcaster}.
In our simulation, humans poll this broadcaster to know about upcoming events. 

For representing a real life behavior, an event broadcaster needs to address questions like where to create event; how often
to create events; and when to start broadcasting information about an event.
We analyzed tweets from Singapore of one week to get locational and temporal distribution of events using the work done in \cite{ruchi:event} and NLTK \cite{nltk}.

\paragraph{\textbf{Events temporal distribution:}}
\label{events-temporal-distribution}
In \cite{ruchi:event}, authors present a system, ET, to extract events from tweets.
ET's event extraction mechanism is based on fact that if a particular keyword shows a sudden
significant increase in its frequency over time, then it can be inferred that some trend / event involving that keyword has started. 
ET takes in a set of tweets as input and divides them in fixed-length time intervals.
Each tweet is tokenized into words and stop words are trimmed out.
Each pair of remaining consecutive words is taken as a bi-gram and constitute a single keyword.
ET extracts frequent keywords in tweets for each interval.
It trims out those keywords that are frequent in all intervals.
Keywords that show a sudden increase in their frequencies across consecutive time blocks are considered as event representative keywords.
Event representative keywords are grouped together using hierarchical agglomerative clustering technique.
The similarity score between two keywords is defined using common co-occurring textual features and similarity in their appearance patterns.
Each resulting cluster represents one event.
ET detects events that range from small events like car accidents to big events like golden globe awards.

Broadcast about an event in simulation starts at time interval of first occurrence of the cluster of related event representative keywords.
We use NLTK \cite{nltk} library to extract precise time of event from related tweets.
We select the most commonly occurring time range as the start and end time of the event. 

\paragraph{\textbf{Events locational information:}}
\label{events-locational-information}
We use NLTK \cite{nltk} library to extract location information about tweets. 
Given a piece of text, it returns a variety of named entities like PERSON, LOCATION, DATE, etc.
For each event that was extracted; or more precisely for each cluster of event representative keywords;
we select tweets that contained them.
We feed these tweets into NLTK and extract location.
Next, we pick the most commonly occurring location in these tweets as location of our event.
We feed this location to Google's Geocoding API and get equivalent latitude and longitude
\cite{google-geocoding}.

\paragraph{\textbf{Determining intended age group for event:}}
Events are often intended towards a particular age group of humans.
For example, while a science exhibition is mainly organized for students, wedding ceremonies usually involve humans of all ages.
Due to privacy issues, Twitter's public API does not provide access to both tweets and age together.
Hence, in our current implementation, we randomly select an age range for the event and broadcast that as part of event information.

In this paper, our focus is not on approach to determine social events and influences in them.
As solutions and accuracy of solutions increase, these solutions can be incorporated as pluggable modules.
Our focus is that our idea is viable and we can use this for better traffic management.

\begin{table}
\centering
\caption{Distribution of human categories}
\begin{tabular}{c|c} \hline
\textbf{Category}&\textbf{Distribution} \\ \hline
\texttt{Working-professional} & 40\% \\
\texttt{Student} & 30\% \\
\texttt{Home-maker} & 15\% \\
\texttt{Senior-citizen} & 15\% \\ \hline
\end{tabular}
\label{distribution}
\end{table}

\section{Humans and their social network}
\label{humans-social-network}
Humans are modeled as proactive agents.
We divide humans into four different categories: working-professional, student, home-maker and senior-citizen; distribution given in Table \ref{distribution}.
Humans makes different type of trips based on their categories as summarized in Table \ref{trips-rules}.

\begin{table}
\centering
\caption{Trips based on human categories}
\begin{tabular}{c|c|c} \hline
  \textbf{Category}             &\textbf{Trip}      &\textbf{Trip starts between} \\ \hline
  Working-professional & home-office       & 7:30-9:30    \\
                                & office-restaurant & 12:00-13:00  \\
                                & restaurant-office & 12:30-13:30  \\
                                & office-home       & 17:30-20:00  \\ \hline
  Student              & home-school       & 7:00-8:00    \\
                                & school-home       & 13:30-14:30  \\
                                & home-other & 18:30-20:30  \\
                                & other-home & 18:30-20:30  \\ \hline
  Home-maker           & home-shop         & 9:00-11:00    \\
                                & shop-home         & 9:30-11:30  \\
                                & home-shop/none    & 18:00-20:00  \\
                                & shop-home         & 18:30-21:00  \\ \hline
  Senior citizen       & home-other/none & 7:00-9:00    \\
                                & other-home   & 8:30-10:00  \\
                                & home-other/none & 17:00-18:30  \\
                                & other-home   & 17:30-19:30  \\ \hline
\end{tabular}
\label{trips-rules}
\end{table}

For example, \textit{senior citizen, home-other/none, 7:00-9:00} indicates that between 7 to 9, a senior citizen will either go from home to some other place or not leave home at all.

We use OSRM (Open Source Routing Machine) \cite{osrm} - a high performance routing engine - for computing shortest routes in road network.
It is incorporated in simulation as a separate service. Each human queries it to find its route from its source to train station and from train station to destination.
We keep maximum speed of human on roads as 35 km/hr - this is to incorporate the fact that humans can use a cab or bus to reach train stations at proper time.

\subsection{Social network of humans}
\label{socialGraph}
\begin{figure}
\centering
\includegraphics[trim=3.8cm 0 3.3cm 0, clip, width=8.8cm, height=5.5cm]{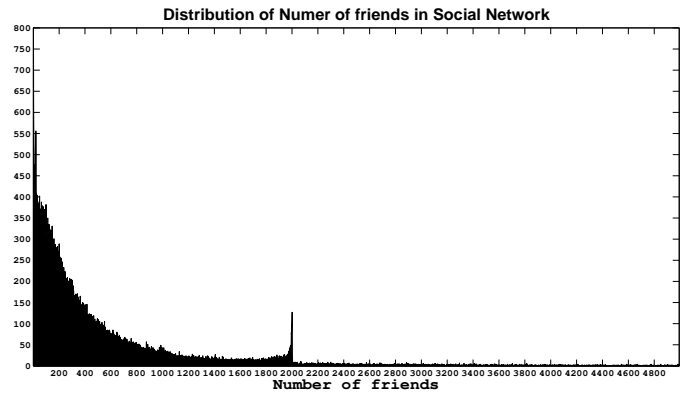}
\caption{Distribution of friends from Twitter dataset}
\label{friends}
\end{figure}

Social network consists of humans and relationships between them.
It is modeled as a directed graph - nodes indicate humans and edge indicates connections(\textit x ``follows'' \textit y) between humans.
Each human in our simulation is a part of social network.
Distribution for number of friends of a human is taken from Twitter dataset
\cite{twitter:dataset} which contains a network of 100,000 humans;
minimum, maximum and average number of friends are 1, 5000 and 500 respectively.
We find that distribution of number of friends decreases exponentially as shown in Figure \ref{friends}.
Our simulation models 100,000 humans and we use above distribution for number of friends. \\

\subsection{Influence Model}
\label{influenceModel}
Influence probability is the likelihood of a human influencing interest of its connection towards a social event.
We consider following factors while computing influence probability.

\subsubsection{Age similarity in human:} A human is likely to influence connections of similar age as there 
is a greater chance that they share similar interests.
Also on several occasions, events are oriented towards a certain age group of humans.
Groups of age and their distribution are presented in Table \ref{age-group-distribution}.
We compute similarity in age of two humans A and B using following formula (where age\_group is computed using Table \ref{age-group-distribution}):
\begin{equation}
\begin{split}
\text{Similar}& \text{AgeInfluence(A, B)} = \\
 & 1 - (|\text{age\_group(A)} - \text{age\_group(B)}|)/6
\end{split}
\end{equation}

\begin{table}
  \centering
  \caption{Age Groups}
  \begin{tabular}{c|c|c}
    \hline
     Age interval & Group No. & Population percentage \\
    \hline
     0-14         & 1         & 13.6     \\
    15-24         & 2         & 18.2     \\
    25-40         & 3         & 25.1     \\
    41-54         & 4         & 25.0     \\
    55-64         & 5         & 09.9     \\
 $\geq$65         & 6         & 08.1     \\
    \hline
  \end{tabular}
  \label{age-group-distribution}
\end{table}
\subsubsection{Similarity in class of human:}
Many times, events are intended for a specific class of humans like conferences for professionals, science exhibitions for students.
So we can say that a student is likely to influence another student more than influencing a working professional for going to
a science exhibition.
Re-using human classes in Table \ref{distribution}, we define similarity between two humans A and B as:
\begin{equation}
\begin{split}
\text{SimilarClass} & \text{Influence(A, B)} \\
 & = 1 \quad \textrm{if (classA == classB)} \\ 
 & = 0 \quad \textrm{otherwise}
\end{split}
\end{equation}
\subsubsection{Proximity in points of contact (homes / offices / schools):} A human is likely
to influence connections that are part of its family, office co-workers, neighbours, go to same school.
Also, if the connection lives nearby, it could get an easy company to visit the event.
Hence, influence would be proportional to nearness in points of contact.
We compute proximity of two people A and B as:
\begin{equation}
\begin{split}
\text{Proximity(A, B)} = \text{Minimum}(& \text{Distance(homeA, homeB),}  \\ 
                            &\text{Distance(officeA, officeB),}  \\ 
                            &\text{Distance(schoolA, schoolB)})  
\end{split}
\end{equation}
If school is undefined for A or B, then we consider distance between their schools as infinite; similarly for offices.
Since A and B can have a different set of connections, influence of A on B is not same as of B on A.
So, to normalize proximity-based influence of A on its connection B,
we divide Proximity(A, B) with the least proximate connection of B.
Hence, we compute proximity-based influence as follows:
\begin{equation}
\begin{split}
\text{Lea}& \text{stProximateConnection(B)} =  \\
&\text{Maximum}(\forall i \in \text{(connections of B), proximity(i, B)})
\end{split}
\end{equation}
\begin{equation}
\begin{split}
\text{Prox} & \text{imityInfluence(A, B)} = \\
& 1 - \text{Proximity(A,B)} / \text{LeastProximateConnection(B)}
\end{split}
\end{equation}

Lastly, we combine above factors with equal weights and say that probability
with which A will successfully influence B towards an event as:
\begin{equation}
\begin{split}
\text{Influence}&\text{Probability(A, B)} = (\\
  &\text{SimilarAgeInfluence(A, B)}   \\
  & + \text{SimilarClassInfluence(A, B)} \\
  & + \text{ProximityInfluence(A, B)}  \\
  &) / 3.0
\end{split}
\end{equation}
We could have also made influence probability a constant for the network, i.e. each human influences all connections by same probability.
We present comparison in results section \ref{influence-probability-model-validation} - based on these results, we found that above model mimics reality more closely than constant influence probability.

\subsection{Human decision to attend an event}
On being interested to attend an event, a human may decide whether to attend it or not.
In real life this could be an outcome of many factors like distance of event from home,
financial cost of attending event, other clashing events, family obligations.
In our simulation we only consider distance factor.
Human estimates time to reach the event based on trains' schedule information available from transport manager (see section \ref{transport-manager}).
A human considers to be "on time" if it can reach to the event within $\pm\tau$ minutes of the start-time, where $\tau$ is 1/10th of duration of event.
If human finds that it can reach to the event on time then it confirms its decision to attend it otherwise there will be no way to attend the 
event even if they wanted.

\subsection{Event information injection in social network}
\label{event-injection}
In day-to-day life, humans check for new interesting information on internet several times a day;
also they do it after certain time intervals and not check it continuously.
In the simulation, humans poll EventBroadcaster (section \ref{event-broadcaster}) with certain probability at
regular intervals to get details about upcoming events.
Upon receiving list of events, a human selects those events which matches its age-group.
It further selects those events where it can reach before event's start time; and tweets about selected events.

\subsection{Information Diffusion using influence probability model}
We now present how information about an event is spread among the connections in social network
and how the interest towards it is aroused in a human using influence probability model defined in Section \ref{influenceModel}.
Intuitively, interest of a human towards an event increases as more and more of its connections in social network present
their desire to attend the event through their posts. We try to capture this motivation.
We use independent cascade model studied in \cite{model:influence}\cite{dkempe-influence} for spreading of interest towards an event in the social network. 

As mentioned earlier, we model social network as a graph where nodes represent humans and edges represent connections.
Whenever a human affirms to attend an event and posts about it, we say that the corresponding node in the graph is active for that event.
Initially, human nodes who injected this information (section \ref{event-injection}) are marked active
and rest of the nodes as inactive with respect to this event.
When a node becomes active, it tries to influence and activate its inactive neighbors (friends) towards the event.
It succeeds with certain probability which we call influence probability (section \ref{influenceModel}) and is distinct for every edge.
On successful activation, neighbor further propagates event information.
Hence, each active node tries to activate its neighbors and the process continues in
discrete steps till no new node is activated, i.e., there is no node which changed from being inactive to active in the current step.

\section{Transport System}
\label{transport-system}
In this section, we describe components that make up transport system.
We use Singapore's metro train network as an example to describe how we select parameters.

\subsection{Station Master}
In real life, a train station needs to be managed with respect to how and when trains can arrive.
Also, during certain time periods ridership of trains may increase and if the trains are not frequent
then it can cause stations to get congested.
The station management may ask the city transport system to provide appropriate measures
like start new trains, increase capacity of existing trains, etc. 
In order to know the state of each train station and communicate it with city transport system,
we appoint a station master agent in each station.
\subsubsection{Tracking ridership at station:}
    Framework implements a simple mechanism for tracking number of humans present at a station at a given time.
    When a human arrives at a station (either from a train or from outside), it tells station master its
    destination station and asks for a token. When human leaves station or boards a train,
    it returns the token to station master. Thus, the number of tokens that are not returned gives
    number of people present at the station. 
\subsubsection{Deciding train arrival at station:}
    Every station has a fixed number of platforms associated with it. So at a given time only a
    limited number of trains can arrive and halt at the station.
    Station master keeps a count of trains that are currently standing at the station.
    When a train is about to arrive, it first asks station master if there is a platform available for it; if not, then train halts at its position. 
    Station master keeps track of all trains waiting to arrive. 
    Whenever a platform is free, station master picks that train which has halted for the longest time.

\subsection{Transport manager}
\label{transport-manager}
In real life, every transport system is managed by a committee which lays out plans for train-routes,
decides schedules and frequency of trains, etc.
In our simulation, we capture this notion through transport manager.
Transport manager decides schedule and capacity of trains to meet changes in ridership of trains. 
Its main motive is to reduce the amount of time humans wait for trains at stations at as less as possible. 
It also provides train inquiry system (section \ref{train-inquiry-system}) with most recent schedule of trains.

\subsubsection{Estimating ridership:}
In order to be able to frame good strategies for schedule and route of trains, the transport manager needs to have
a good estimation of ridership at various stations during various hours of the day.
Transport manager actively monitors ridership at various stations by communicating with station masters and keeps track of it for future estimations.
It also tries to estimate ridership based on upcoming social events.
It continuously polls event broadcaster (section \ref{event-broadcaster}) about schedule and location of upcoming events.
For each event, it extracts data from social network about humans that are planning to visit.
Based on humans' locations and the time of event it estimates change in ridership of train routes using procedure in Table \ref{change-in-ridership}.
Sum of regular ridership (as obtained from station masters) and the estimated change due to events gives final ridership. 
\begin{table*}[ht] \footnotesize
\centering
\begin{tabular}{l}
  \hline
  \texttt{Construct a directed graph with train stations as nodes and train routes as edges.} \\[0.5ex]
  \hspace{2mm} \texttt{for every hour of next 24 hours:} \\ [0.5ex]
  \hspace{4mm} \texttt{for every event in this hour:} \\ [0.5ex]
  \hspace{6mm} \texttt{Find location of the event.} \\ [0.5ex]
  \hspace{6mm} \texttt{Find train station nearest to event location and map it on the graph as destination. } \\ [0.5ex]
  \hspace{6mm} \texttt{for each human attending this event:} \\ [0.5ex]
  \hspace{8mm} \texttt{Estimate whether human will come from home / school / office based on its category} \\
  \hspace{8mm} \texttt{and time of day.}\\ [0.5ex]
  \hspace{8mm} \texttt{Find location of human based on above estimation.} \\ [0.5ex]
  \hspace{8mm} \texttt{Find train station nearest to human location and map it on the graph as source for} \\
  \hspace{8mm} \texttt{above destination.} \\  [0.5ex]
  \hspace{4mm} \texttt{for each train route:} \\ [0.5ex]
  \hspace{6mm} \texttt{Traverse the route and count number of sources that come before their destination.}\\ [0.5ex]
  \hspace{6mm} \texttt{Mark above count as change in ridership of this train route for this hour.} \\ [0.5ex]
  \hline
\end{tabular}
\caption{Procedure to estimate change in ridership}
\label{change-in-ridership}
\end{table*}

\subsubsection{Capacity of trains:}
Capacity of trains is defined as number of seats in it.
Initially, each train has same capacity; transport manager may strategize to change it as and when needed.
Defining strategies to change capacity is a plug-and-play module for users of our framework.
An important feature of our simulated population is that each human rides 
trains at least once a day and most humans rides multiple times - this is not true
in real-life where a significant number of people do not ride trains or use trains less often.
Hence, capacity of trains is chosen based on ratio of ridership of Singapore's metro trains in real-life to ridership in our simulation.  
From \cite{slta-stats}, the daily ridership is 2.3 million; and each train has capacity of 1920 (\cite{siemens-c561}).
In our simulation, we have a ridership of 370,000 on a regular day.
Hence, we compute initial capacity of trains as:

    $370,000 * 1920 / 2,300,000 = 310$

\subsubsection{Schedule of trains:}
We initialize train schedule using Singapore's metro schedule taken from \cite{singapore:metro-website}.
Again strategizing and testing schedules is a pluggable feature of our framework.
In order to evaluate efficiency of a schedule (or capacity), our framework computes two metrics - train-usage and average waiting-time - and presents its results to the users.

\paragraph{Train usage:}
Schedules should be economical; for example, one may decide to increase frequency of trains on certain routes in order to reduce waiting times; but this will lead to increase in cost of running the trains and there may be a large number of seats that stay unoccupied
for large part of train trips.
Train usage expresses above rationale. For a given time interval $T$, it is defined as: 

$U_T = (\sum_{i} t_{iT}) / (C * T)$

where $t_{iT}$ is total time for which seat $i$ was occupied (need not be for contiguous duration) in time interval $T$,
and $C$ is capacity of the train.

\paragraph{Average waiting time:}
Average waiting time of humans for given time interval $T$ is defined as:

$W_T = \sum_{i} t_{iT} / N$

where $t_{iT}$ is total time for which a human waited for train at any station during time interval $T$ and $N$ is total number of humans.

\subsection{Train inquiry system}
\label{train-inquiry-system}
In real life, it is a common practice that humans inquire about train schedule on phones, metro-websites, 
on bulletin boards at train stations, etc.
We encapsulate all sources of train related information into a single module which we call train inquiry system.
Train inquiry system continuously polls transport manager for getting most updated schedule of trains.
It also continuously polls station masters to know about current location and delays of trains.
Humans query this system for train information and plan their trips accordingly.

\begin{figure*}
        \centering
        \begin{subfigure}[lt]{0.5\textwidth}
                \centering
                \includegraphics[trim= 3.1cm 0 3.1cm 0, clip, width=\textwidth, height= 5cm]{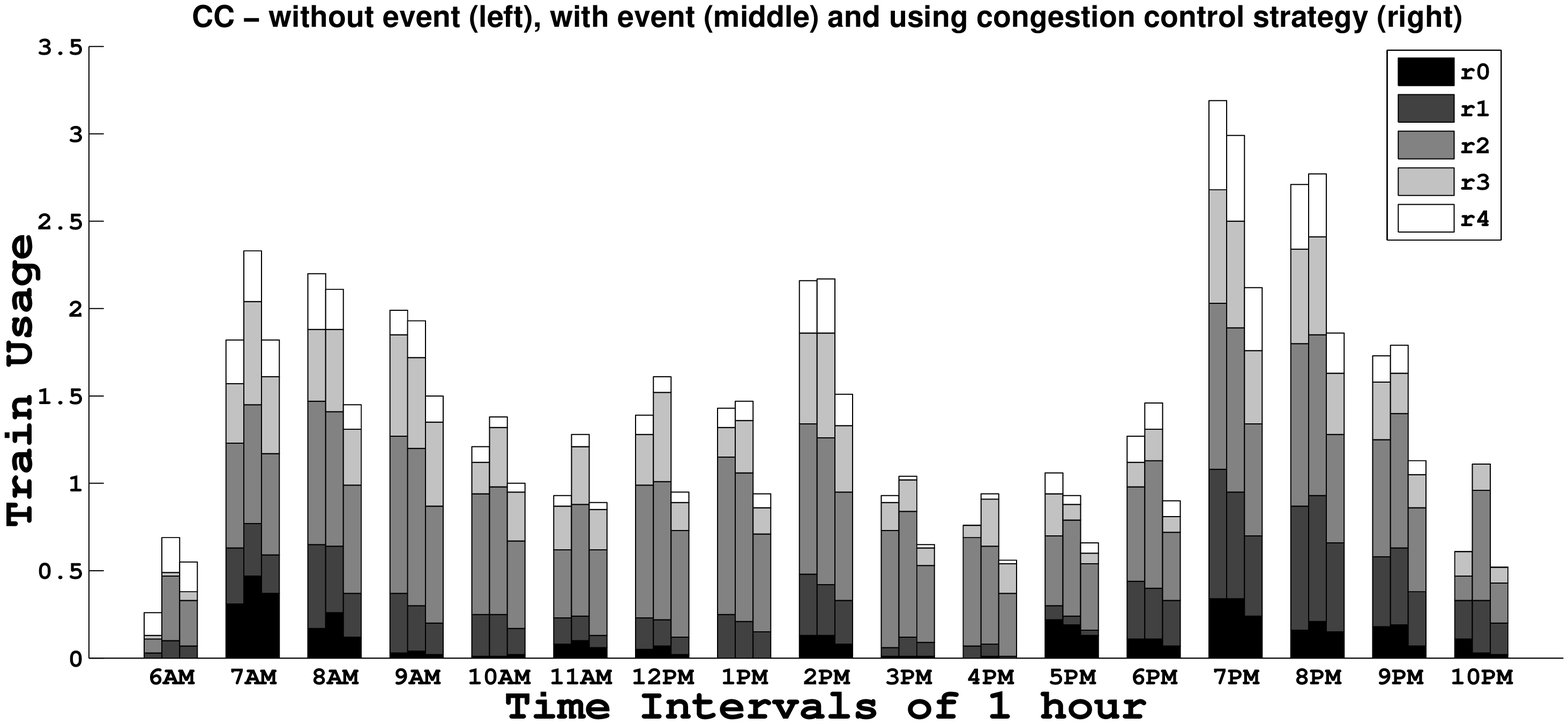}
                \label{usage:CC}
        \end{subfigure}%
        ~ 
        \begin{subfigure}[tl]{0.5\textwidth}
                \centering
                \includegraphics[trim= 3.1cm 0 3.2cm 0, clip, width=\textwidth, height=5cm]{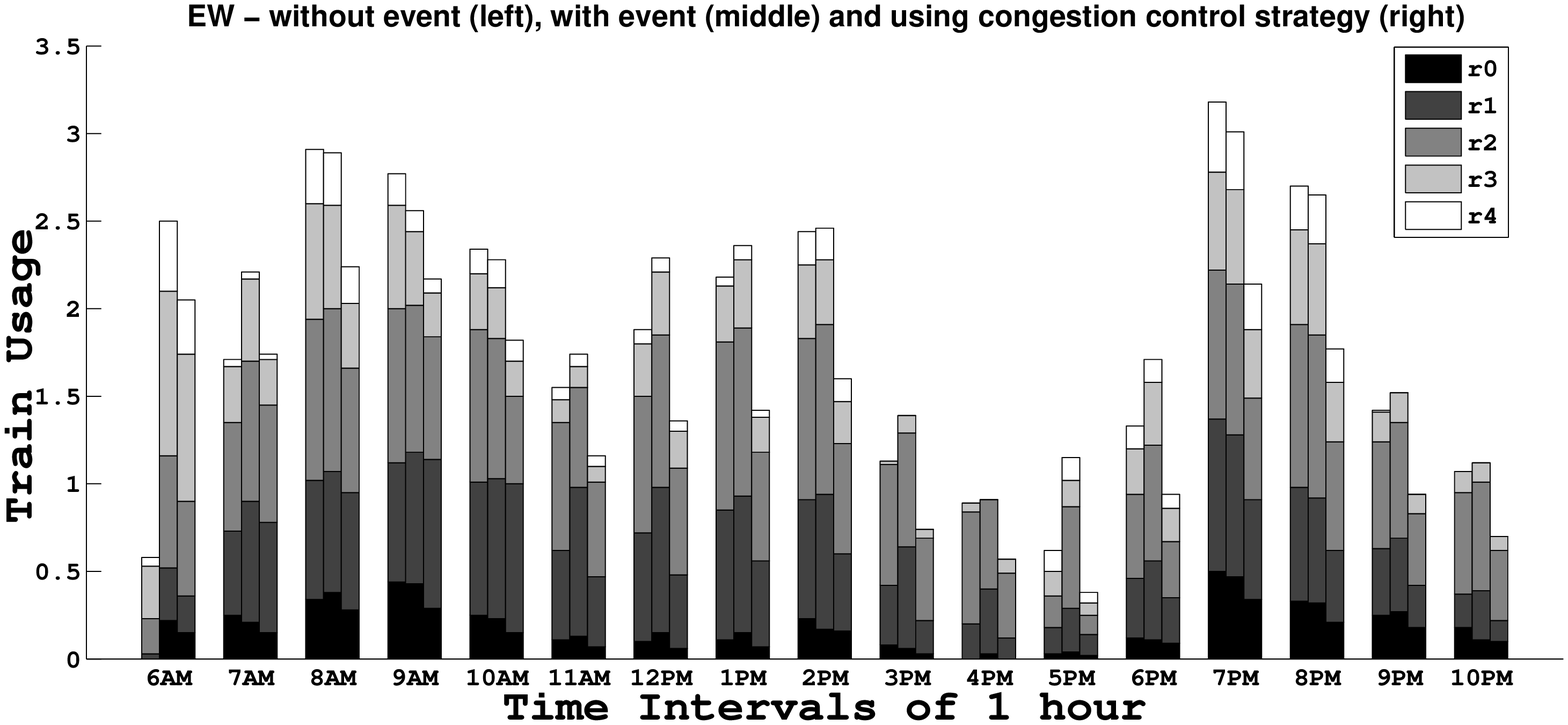}
                \label{usage:EW}
        \end{subfigure}
\setlength{\abovecaptionskip}{-2pt}
\setlength{\belowcaptionskip}{-2pt}
        \caption{Comparison of Train Usage across CC and EW train lines}
        \label{usage-comparison}
\end{figure*}

\section{Results}
\label{results}
We simulate train transport system of Singapore with a population of 100,000 humans.
We simulate four major metro lines of Singapore rail network - North South Line (NS), North East Line (NE), East West Line (EW) and Circular Line (CC) with a total of 87 train stations (\cite{singapore:metro-website}). 
Trains have capacity for 310 (section \ref{transport-manager}) and follow real-life schedule.

We divide results on validity of framework design and parameters based on following aspects.
\begin{figure}
\setlength{\abovecaptionskip}{-2pt}
\setlength{\belowcaptionskip}{-2pt}
\centering
\includegraphics[trim= 2.5cm 0 1.6cm 0, clip, width=9cm, height=5.5cm]{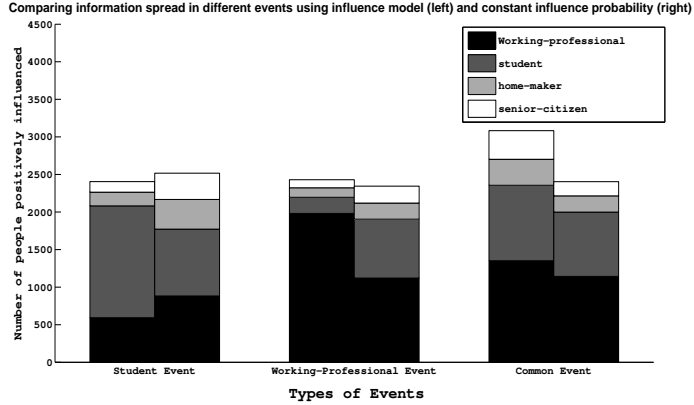}
\caption{Comparing humans influenced positively for events}
\label{influenceStack}
\end{figure}

\subsection{Influence probability model}
\label{influence-probability-model-validation}
We simulated three events: a science exhibition, a professional conference and a wedding event and studied the distribution
of human in terms of both age and class that were influenced positively towards these events.
Results showed that using our influence probability model (section \ref{influenceModel}),
ratio of students that attended science exhibition was much higher than other class of humans;
similar results were obtained for age-groups (see Fig. \ref{influenceStack}).
Next, we made influence probability constant for each connection and set it to 0.5.
Using constant probability model, we get almost same ratio of humans,
both in terms of class and age groups, that attended science exhibition.
On the other hand, for wedding we see a similar distribution of various age-groups attending the event in both cases.
Figure \ref{influenceStack} shows the comparisons.
Hence, we see that per-connection varied influence probability model mimics reality closely and performs better than constant influence probability in this aspect.

\begin{figure*}
        \centering
        \begin{subfigure}[lt]{0.5\textwidth}
                \centering
                \includegraphics[trim= 3.2cm 0 3.1cm 0, clip, width=\textwidth, height= 5cm]{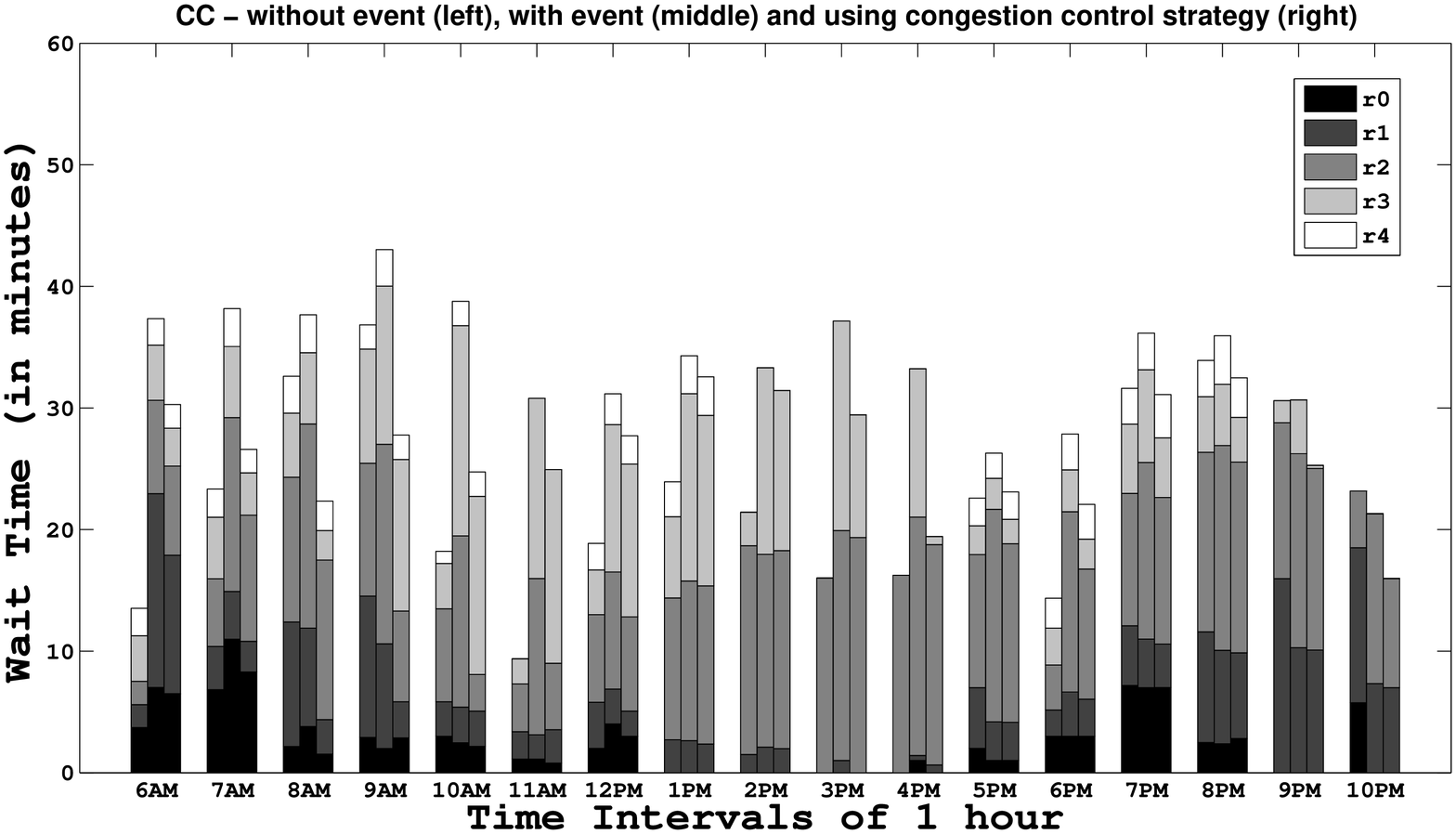}
                \label{wait:CC}
        \end{subfigure}%
        ~ 
        \begin{subfigure}[tl]{0.5\textwidth}
                \centering
                \includegraphics[trim= 3.2cm 0 3.2cm 0, clip, width=\textwidth, height=5cm]{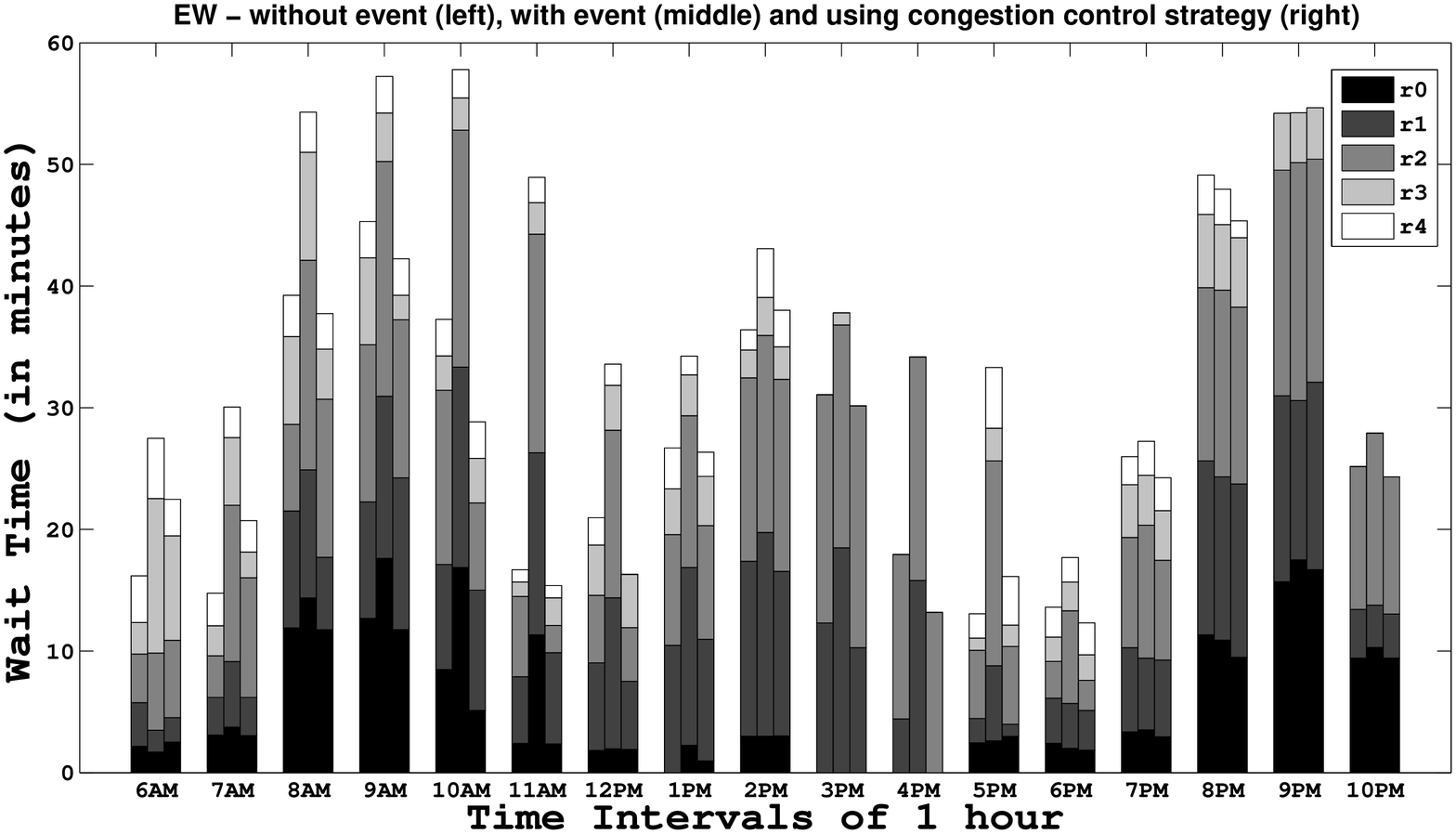}
                \label{wait:EW}
        \end{subfigure}
\setlength{\abovecaptionskip}{-2pt}
\setlength{\belowcaptionskip}{-2pt}
        \caption{Comparison of average waiting time across CC and EW train lines}
        \label{wait-comparison}
\end{figure*}

\subsection{Effect of events}
We ran two simulations - with and without a social event.
Event was held near stations EW21 and CC22 between 8:30 to 11:30 am.
In order to get fine-grained look at the usage and wait-time metrics, we divide each train-line into 5 sections - \textit{r0, r1, r2, r3, r4} - each section has equal number of consecutive stations.
For a given time interval, usage of a section of train line is summation of usage of trains that cross this section.
Similarly, for a given time interval, waiting time of a section of train line is average of wait times for humans standing on stations in that section of train line.
Comparison of usage of trains and average wait-time are shown in figures \ref{usage-comparison}, \ref{wait-comparison}.
We see an increased usage on EW and CC lines during event hours.
NE and NS lines also see a small increase but results are omitted due to space limitations.
Also, we see that train usage is higher in middle sections of lines rather than tail stations (irrespective of events) which matches real-life behavior.

\subsection{Effect of congestion control strategies}
We implemented a greedy congestion control strategy as follows.
Transport manager monitored upcoming events and estimated riderships at the beginning of every hour.
If it saw that ridership increased beyond capacity of trains, then it increased capacity of train by adding a compartment.
Transport manager has only a fixed number of compartments and each compartment seats 31 humans.
So, if transport manager is out of compartments, it detaches the compartment from a train whose ridership has decreased.
It also assigns compartment in order of decreasing ridership of affected trains.
With this strategy, Figure \ref{wait-comparison} shows that wait-time was reduced significantly (as compared to fixed capacity model). 
Though the effective train usage was slightly decreased \ref{usage-comparison} as transport manager detaches compartment only when it needs.

\subsection{Strategies involving human behavior}
We implemented a strategy in which human alters its trip based on state of ridership in trains.
Initially, a human selects route which takes shortest total travel time to reach destination.
Human keeps polling station master to get the current state of ridership of trains.
If a human perceives that train involved in its route is full, then it computes alternative routes to destination and selects one with next minimum total travel time. 
Total travel time is defined as sum of time spent in traveling from source to destination including time taken to go to and fro train stations and wait time at station.
Waiting at the same train station for next train running on original route is also considered a valid alternate route.
We incorporated human behavior exercising alternative routing in simulation and found that 10\% of humans choose to take alternates.
Table \ref{alternate-routes-table} shows comparison of average waiting time and average total trip time 
in experiment with or without using alternative routing approach.
It is noted that there is slight difference in both of metrics.

\begin{table}
  \centering
  \caption{Alternative Routing Approach}
  \begin{tabular}{l|c|c} \hline
    &\textbf{Without}&\textbf{With}\\ 
    &\textbf{alternate routing}&\textbf{alternate routing}\\ \hline
    \textbf{Average waiting time}  & 34.3 mins & 31.7 mins\\ 
    \textbf{Average total travel time} & 82.8 mins & 79.5 mins\\ \hline
  \end{tabular}
  \label{alternate-routes-table}
\end{table}

\section{Conclusion}
\label{conclusion}
Traffic congestion wastes time and money; and needs to be controlled effectively.
Social networks can provide ample information about upcoming events and locational distribution of people planning to attend such events.
We presented basic building blocks of a framework that simulates train transport system of a city, human behavior and a social network connecting them; and aims to capture real-life like behavior.
It provides researchers a platform to further experiment in this area easily.
Future work includes incorporating buses and cars as well as modes of transportation; and documenting and open-sourcing the framework for public use.


%
%
\bibliographystyle{abbrv}
\bibliography{paper}  
%
\end{document}